\documentclass[doublecol]{epl2}

\usepackage{amsmath}
\usepackage{amssymb}
\usepackage{graphicx}
\usepackage{bm}
\usepackage{epstopdf}

\newcommand{\prt}{\partial}
\newcommand{\sn}{\mathrm{sn}}
\newcommand{\ga}{\gamma}
\newcommand{\Ga}{\Gamma}

\newcommand{\Om}{\Omega}
\newcommand{\la}{\lambda}
\newcommand{\K}{\mathrm{K}}

\title{Quasi-one-dimensional flow of polariton condensate past an obstacle}
\shorttitle{Flow of polariton condensate past an obstacle}

\author{Anatoly M. Kamchatnov\inst{1}\thanks{E-mail: \email{kamch@isan.troitsk.ru}}
\and Yaroslav V. Kartashov\inst{2}\thanks{E-mail: \email{yaroslav.kartashov@icfo.es}}}
\shortauthor{A. M. Kamchatnov \etal}

\institute{
\inst{1}  Institute of Spectroscopy, Russian Academy of Sciences, Troitsk,
Moscow Region, 142190, Russia\\
\inst{2} ICFO-Institut de Ciencies Fotoniques, and Universitat Politecnica de Catalunya,
Mediterranean Technology Park, 08860 Castelldefels (Barcelona), Spain
}

\abstract{
Nonlinear wave patterns generated by the flow of polariton condensate past an obstacle are studied for quasi-one-dimensional
microcavity geometry. It is shown that pumping and nonlinear damping play a crucial role in this process leading to sharp
differences in subsonic and supersonic regimes. Subsonic flows result in a smooth disturbance of the equilibrium condensate
around the obstacle whereas supersonic flow generates a dispersive shock wave in the flow upstream the obstacle and a long
smooth downstream tail. Main characteristics of the wave pattern are calculated analytically and analytical results are
in excellent agreement with the results of numerical simulations. The conditions for existence of stationary wave patterns are
determined numerically.}

\begin{document}

\maketitle

\section{Introduction}
Recent experimental progresses in studying the microcavity polaritons have lead to
a huge growth of interest in their collective dynamics
(see, e.g., review articles~\cite{keeling-2007,amo-2010} and references therein).
Polaritons possess an extremely small effective mass which allows their
condensation at temperatures much greater than that of ultracold atomic vapors.
Besides that, parameters of the polariton superfluid can be easily tuned with
the use of resonant lasers. However,
polaritons have a finite lifetime, and to maintain their steady-state population a
continuous pumping is required. Experimentally, it is observed
\cite{kasprzak-2006,lai-2007} that above a threshold pumping strength
an accumulation of low
energy polaritons is accompanied by a significant increase of spatial coherence that extends over the entire cloud of
polaritons which can then be described by a single order parameter (polariton condensate wave function)
obeying an effective Gross-Pitaevskii equation.
On the contrary to the atomic condensate situation, the density of the polariton condensate is not an arbitrary parameter
anymore. Instead, it is determined by the condition of balance between pumping and dissipation processes. This implies that
the non-conservative effects can play a crucial role in the condensate's nonlinear dynamics. For example, the
generation of oblique solitons by the flow of a condensate past a localized obstacle has been observed~\cite{amo-2011} at
subsonic speed which is impossible in the case of conservative atomic condensate. The formation of oblique solitons followed
by their decay into vortex streets in a non-uniform cloud of polariton condensate has been
studied in~\cite{grosso-2011}. Other geometries are also of great interest. In particular, quasi-1D flow of polariton
condensate along ``quantum wires'' was studied experimentally in \cite{wertz-2010}. In the atomic condensate case
such a flow past an obstacle leads to generation of dispersive shock waves propagating upstream and downstream from the
obstacle \cite{hakim-1997,pavloff-2002,radouani-2004}, as it was observed in the experiment~\cite{ea-2007} and explained theoretically in~\cite{legk}. However, it is
easy to see that this theory cannot be applied to the polariton condensate flow which density must be fixed (if
pumping and dissipation are balanced) far enough from the obstacle thus preventing formation of jumps in the flow parameters.
This means that in the dissipative case the dispersive shock waves generated by the flow must always be
attached to the obstacle
and relax to the steady-state flow far enough from the obstacle. This qualitative difference between wave patterns in
conservative and dissipative cases makes the theory of dissipative flow much more complicated and this Letter is devoted
to its development.

\section{Theoretical model}
Several models have been suggested for theoretical description of polariton condensate
(see, e.g., \cite{km-2003}). They were based on various generalizations of the
Gross-Pitaevskii (GP) equation. Here we have to take into account the effects of losses and
pumping of polaritons on generation of dispersive shock waves by the flow of condensate
past an obstacle. To this end, we will use the simple model introduced in
\cite{kb-2008} where nonresonant pumping due to stimulated scattering of
polaritons into the condensate and their linear losses are described by the effective
``gain'' term $\prt_t\psi=\ga\psi$ where $\psi$ is the polariton condensate wave
function. The saturation of gain is modeled by the nonlinear term
$\prt_t\psi=-\Ga|\psi|^2\psi$ which brings the condensate density into
equilibrium with the one of external reservoir, $\rho\equiv|\psi|^2=\ga/\Ga$.
The dispersion of the lower branch of polaritons in the
effective mass approximation and the nonlinear interaction due to the
exciton component  of polaritons lead to the following generalized GP equation
\begin{equation}\label{eq1}
    i\psi_t+\tfrac12\psi_{xx}-|\psi|^2\psi=V(x)\psi+i(\ga-\Ga|\psi|^2)\psi,
\end{equation}
(written in standard non-dimensional units) where $V(x)$ denotes the potential
of the obstacle
and $x$ is the direction of the flow of the condensate. The theory developed below
can be generalized to other forms of nonlinear gain as, for
example, the model considered in~\cite{wc-2007},
and the results remain qualitatively similar. However, to be definite, we shall consider here
the model (\ref{eq1}) for which the description looks especially simple.

\begin{figure}
\centerline{\includegraphics[scale=0.11]{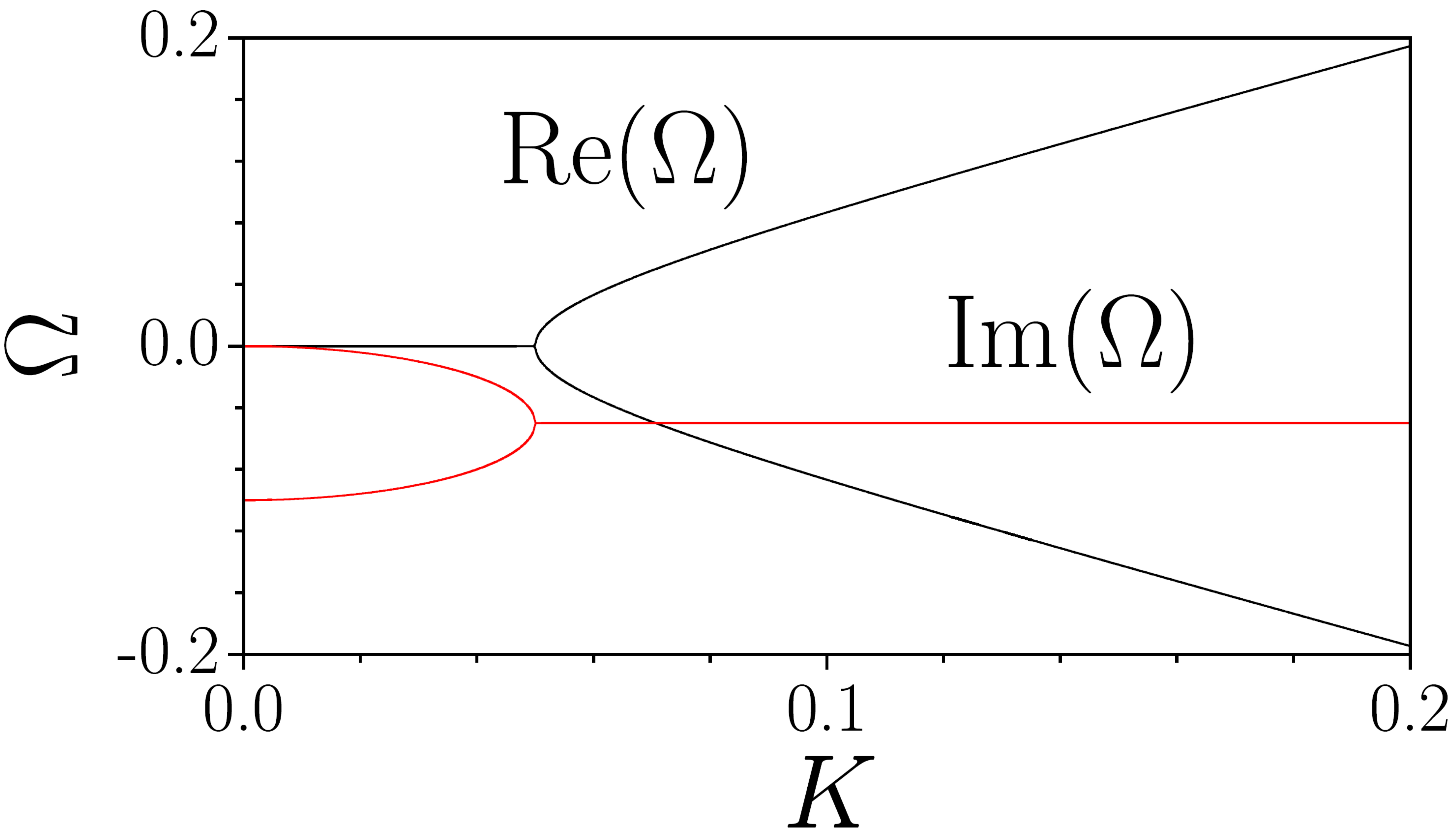}}
\caption{(Color online) The real and imaginary parts of $\Omega$
as functions of $K$ (see eq.~(\ref{eq3})) are shown in the case  $\ga=\Ga=0.05$  and  $u_0=0$. }
\label{fig.1}
\end{figure}

It is easy to see that eq.~(\ref{eq1}) without potential ($V(x)\equiv0$) admits a
plane wave solution which constant amplitude is determined by the gain and
damping coefficients,
\begin{equation}\label{eq2}
    \psi=ae^{i(u_0x-\mu t)},\quad a=\sqrt{{\ga}/{\Ga}},\quad
    \mu={\ga}/{\Ga}+\tfrac12u_0^2,
\end{equation}
where $u_0$ denotes the uniform flow velocity of the condensate and
$\rho_0={\ga}/{\Ga}$ is its density.
The analysis of modulation stability of such plane waves with respect to
harmonic perturbations $\propto \exp[i(Kx-\Om t)]$ shows that the disturbance
propagates along the wave (\ref{eq2}) with the dispersion law
\begin{equation}\label{eq3}
 \Om=u_0K-i\ga\pm\sqrt{K^2(a^2+K^2/4)-\ga^2}.
\end{equation}
The expression under the square root vanishes at the wave number
\begin{equation}\label{eq4}
    K_c=a\left[2(\sqrt{1+\Ga^2}-1)\right]^{1/2}
\end{equation}
which separates two different regimes of evolution of a harmonic perturbation;
see the plots in fig.~\ref{fig.1}.
In the conservative limit $\ga,\,\Ga\to0,\,\ga/\Ga=\mathrm{const}$ we reproduce
the standard Bogoliubov dispersion law with sound velocity $c_s=a=\sqrt{\rho_0}$.
We shall call this parameter ``sound velocity'' also for small
$\ga,\,\Ga\ll1$; it corresponds to the almost linear part $\Om\approx c_sK$
of the dispersion law for $K_c\approx \Ga a \ll K \ll a$ (see fig.~\ref{fig.1}).
Notice that, as long as repulsive nonlinearity is considered, the plane wave solution
(\ref{eq2}) is modulationally stable, since $\mathrm{Im}(\Omega)<0$  for any $K$-value.

It is convenient to transform eq.~(\ref{eq1}) into a hydrodynamic form by means of the substitution
\begin{equation}\label{eq5}
    \psi=\sqrt{\rho}\,\exp\left(i\int^x u(x',t)\upd x'\right)e^{-i\mu t}
\end{equation}
which yields
\begin{equation}\label{eq6}
    \rho_t+(\rho u)_x=2\Ga\rho(\rho_0 - \rho),
\end{equation}
\begin{equation}\label{eq7}
u_t+uu_x+\rho_x+\left(\frac{\rho_x^2}{8\rho^2}-\frac{\rho_{xx}}{4\rho}\right)_x=-V_x.
\end{equation}
The last term in the left-hand side of (\ref{eq7}) describes the dispersion effects; the right-hand side of
(\ref{eq6}) describes the gain and loss effects in the system; the other terms in
eqs.~(\ref{eq6}),(\ref{eq7}) have standard hydrodynamic meaning.

\section{Hydraulic approximation}
In what follows we assume that the size $l$ of the obstacle is much greater than the
``healing length'' (equal to unity in our non-dimensional variables). In our
numerical simulations we will use an obstacle potential in the form
\begin{equation}\label{eq8}
    V(x)=V_m\exp(-x^2/l^2),\quad l\gg1.
\end{equation}
All numerical plots below are obtained for $l=5$.
We are interested in stationary patterns with
$\rho_t\equiv0,\,u_t\equiv0$ generated by the steady flow of the condensate past a
localized obstacle ($V(x)\to0$ as $|x|\to\infty$), i.e.
$\rho$ and $u$ must satisfy the boundary conditions
\begin{equation}\label{eq9}
    \rho\to \rho_0,\quad u\to u_0\quad \text{as}\quad |x|\to\infty.
\end{equation}
Then eq.~(\ref{eq6}) can be reduced to
\begin{equation}\label{eq10}
    (\rho u)_x=2\Ga\rho(\rho_0-\rho)
\end{equation}
and the stationary eq.~(\ref{eq7}) can be integrated once to give
\begin{equation}\label{eq10a}
    \tfrac12{u^2}+\rho+\frac{\rho_x^2}{8\rho^2}-\frac{\rho_{xx}}{4\rho}+V(x)=
    \tfrac12{u_0^2}+\rho_0
\end{equation}
For $l\gg 1$ it is natural to assume that the wave pattern has the
characteristic length about $l$ and, hence, the dispersive terms in
(\ref{eq10a}) having higher order derivatives are negligibly small compared with
other terms; then we get
\begin{equation}\label{eq11}
    \tfrac12{u^2}+\rho+V(x)=
    \tfrac12{u_0^2}+\rho_0.
\end{equation}
Integration of eq.~(\ref{eq10}) over space shows that stationary patterns
satisfy the condition (see also \cite{kb-2008})
\begin{equation}\label{eq12}
    \int_{-\infty}^{\infty}\rho(\rho_0-\rho)\upd x=0.
\end{equation}
Excluding $u$ from Eqs.~(\ref{eq11}) and (\ref{eq10}) we get the equation
\begin{equation}\label{eq13}
    \left(\rho\sqrt{u_0^2+2(\rho_0-\rho-V(x))}\right)_x=2\Ga\rho(\rho_0-\rho)
\end{equation}
which determines the dependence of $\rho$ on $x$ for a given potential $V(x)$
provided the solution satisfies the condition (\ref{eq9}).
This equation represents the so-called hydraulic approximation for our system.
(It generalizes the well-known Thomas-Fermi approximation to non-zero flow
velocity.)

At the tails of the wave pattern ($|x|\to\infty$) we can neglect the potential $V(x)$
and linearize eq.~(\ref{eq13}) with respect to small deviations $\rho-\rho_0$
from the steady state. This gives the asymptotic behavior
\begin{equation}\label{eq14}
    |\rho-\rho_0|\propto\exp\left(\frac{2\Ga\rho_0u_0}{\rho_0-u_0^2}x\right)
\end{equation}
which shows that such tails extending beyond the range of the potential
can exist only under the condition
\begin{equation}\label{eq15}
    \frac{|\rho_0-u_0^2|}{\Ga\rho_0u_0}\gg l.
\end{equation}
This means that the sound velocity $c_s=\sqrt{\rho_0}$ separates two
different regimes---subsonic ($u_0<c_s$) and supersonic ($u_0>c_s$)---with
drastically different properties. We shall discuss them separately.

\section{Subsonic flow}
\begin{figure}
\centerline{\includegraphics[scale=0.11]{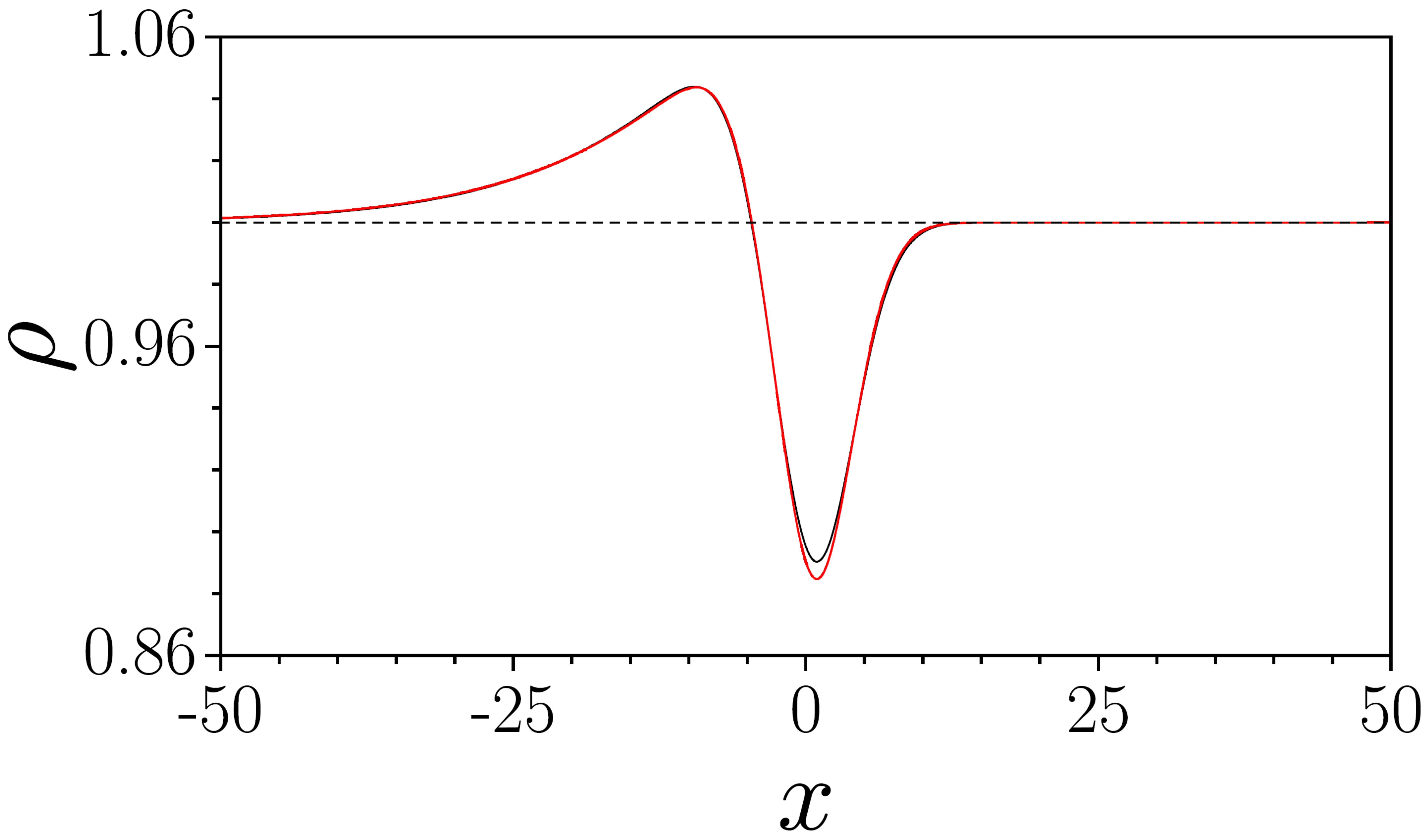}}
\caption{(Color online) Comparison of density profile
at  $u_0=0.6$, $V_m=0.1$, and $\ga=\Ga= 0.05$ obtained numerically from
eq.~(\ref{eq1}) (black curve)
and in the framework of the hydraulic approximation (red curve). The dashed line shows
plane wave solution (\ref{eq2}) in absense of potential. }
\label{fig.2}
\end{figure}

\begin{figure}
\centerline{\includegraphics[scale=0.11]{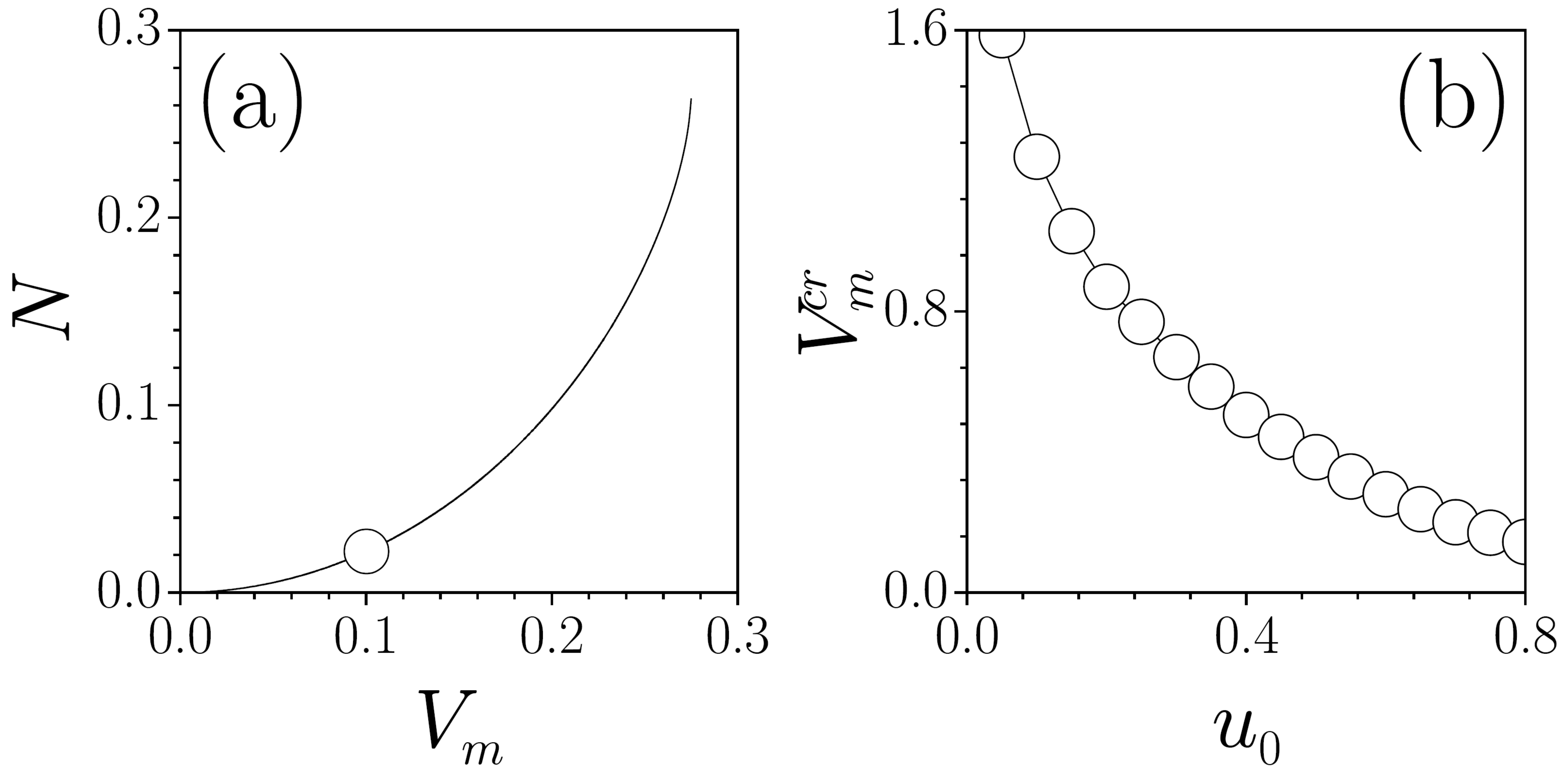}}
\caption{(a) Number of particles (\ref{number}) versus potential strength
for subsonic flow with  $u_0=0.6$. The circle corresponds to the parameters used in fig.~\ref{fig.2}.
(b) Critical strength of potential
versus  $u_0$. In all cases  $\ga=\Ga=0.05$.}
\label{fig.3}
\end{figure}

The hydraulic approximation describes the profiles of the disturbance well enough practically
in the entire subsonic regime $u_0<\sqrt{\rho_0}$ if the velocity is not too close
to the sound velocity. In fig.~\ref{fig.2}
we compare the direct numerical solution of eq.~(\ref{eq1}) with its hydraulic approximation
obtained by the numerical solution of eq.~(\ref{eq13}); quite good agreement is observed
(notice that in all simulations we use $\ga=\Ga=0.05$ that gives
sound velocity $c_s=1$).
Naturally, the asymptotic behavior (\ref{eq14}) is also confirmed in the upstream flow $x<0$. However,
there is no linearized solution decaying to zero at $x\to+\infty$, hence the transition to the
asymptotic steady flow in the downstream region $x>0$ has to occur within the range of the
potential $V(x)$, and this conclusion is confirmed by direct numerical solution.
Indeed, as one observes in fig.~\ref{fig.2}, the stationary disturbance has a smooth shape with
a very long monotonically decaying left tail.
This tail is attached to the density dip located almost completely within the potential region.
The amplitude of the disturbance (i.e. the difference between maximal and minimal density in the
condensate) in the subsonic regime monotonically increases with growth of the potential strength.
It is convenient to introduce the ``number of particles''
\begin{equation}\label{number}
    N=\int_{-\infty}^{\infty}[\rho^{1/2}(x)-a]^2\upd x
\end{equation}
disturbed by the flow.
This variable monotonically increases with the strength of potential [fig.~\ref{fig.3}(a)].
Interestingly, there exists a critical strength
of potential $V_m^{cr}$ above
which one cannot find stationary profiles. When $V_m\to V_m^{cr}$  the tangent
to $N(V_m)$ becomes vertical. For this strength of potential the amplitude of disturbance is maximal.
Note that the critical potential strength $V_m^{cr}$ diverges when the incident velocity of the
condensate $u_0\to0$ and it monotonically decreases when $u_0$ increases
[fig.~\ref{fig.3}(b)]. We were able to obtain the critical potential strength only for
$0<u_0<0.8c_s$. For higher velocities $[0.8c_s<u_0<c_s]$ the condensate starts
developing small oscillations on its right tail. Solutions characterized by different
number of oscillations on their right tail may form continuous families that may be obtained
even for potentials with $V_m>1$.

The condition (\ref{eq15}) breaks down when the incident velocity $u_0$ is close to the critical value
$\sqrt{\rho_0}$. In this case the potential $V(x)$ cannot be neglected in the
hydraulic approximation (\ref{eq13}. Even more, as $u_0$ approaches the critical value, the upstream
density profile steepens and its slope can become so large that the dispersive terms
in eq.~(\ref{eq10a}) can no longer be neglected. As is known, the dispersive effects lead to generation of
oscillations in the regions with fast change of the variables. Hence, for $u_0\gtrsim \sqrt{\rho_0}$
we should expect generation of dispersive shock waves and now we proceed to the discussion
of the supersonic flow.

\section{Supersonic flow}
If $u_0>\sqrt{\rho_0}$ and the condition (\ref{eq15}) is fulfilled, then the linearized solution
(\ref{eq14}) describes decaying disturbance in the downstream flow at $x\to+\infty$. Thus,
contrarily to the subsonic case, the right tail for $x>0$ can extend far beyond the range of
the potential.
\begin{figure}
\centerline{\includegraphics[scale=0.12]{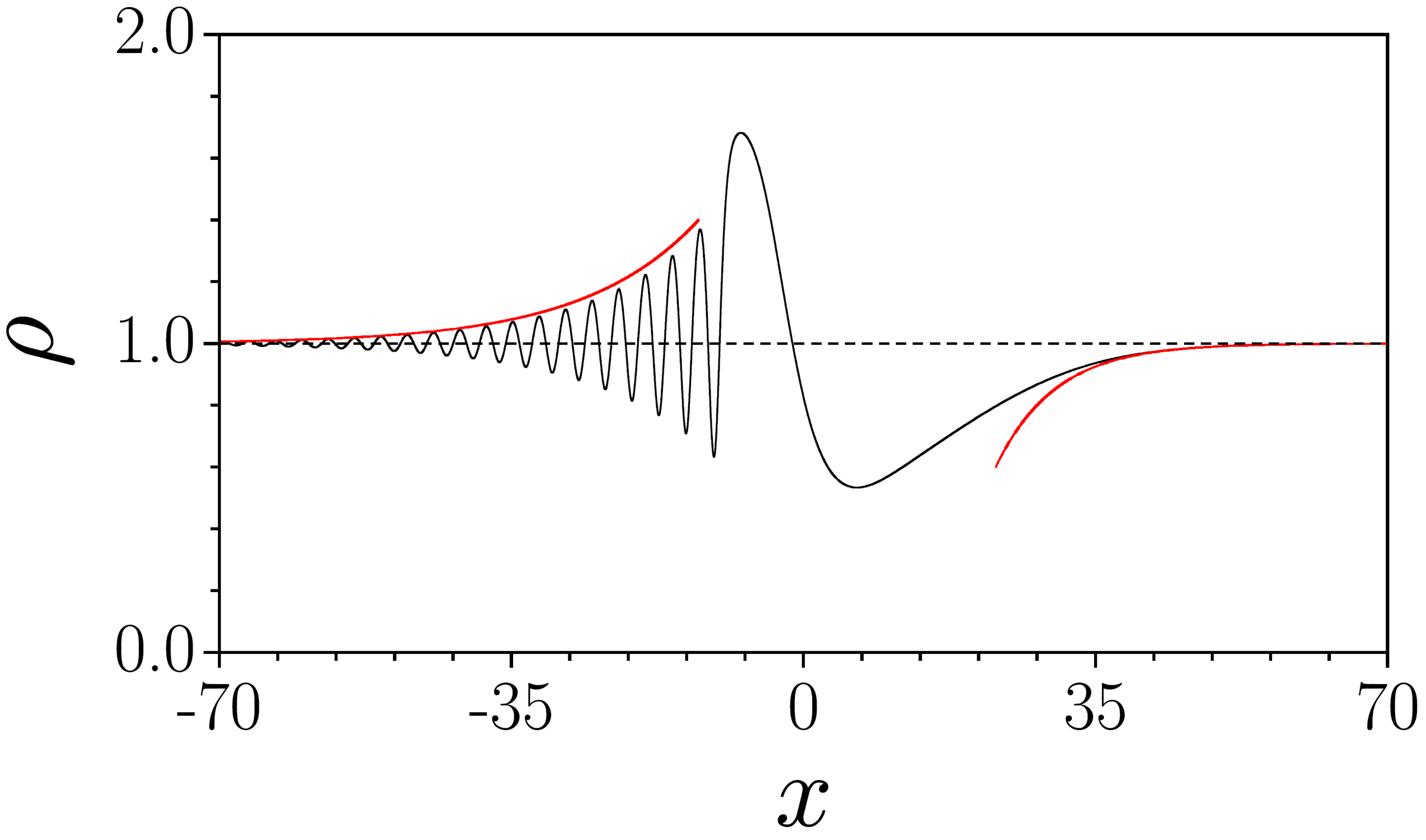}}
\caption{(Color online) The density distribution for shock wave at $u_0=1.4$,
$V_m=0.8$,
and $\ga=\Ga=0.05$. Red curves show analytical approximations of tails from
eqs.~(\ref{eq14}) and (\ref{eq40}). Dashed line shows plane wave solution in the absence of potential. }
\label{fig.4}
\end{figure}

As mentioned above, the generation of an oscillatory dispersive shock wave is expected in the upstream flow.
This is confirmed by numerical solution of (\ref{eq1}) with the supersonic boundary conditions;
see fig.~\ref{fig.4}. The dispersive shock waves can be represented as a modulated solution of the undisturbed ($\Ga=\ga=0,\,V(x)\equiv0$)
equation (\ref{eq1}) or the system (\ref{eq6}), (\ref{eq7}) which can be written in the form (see, e.g., \cite{kamch2000})
\begin{equation}\label{eq16}
    \begin{split}
    \rho(x,t)&=\nu_1+(\nu_2-\nu_1)\sn^2(\sqrt{\nu_3-\nu_1}\,\theta,m)\\
    &=\tfrac14(\la_1-\la_2-\la_3+\la_4)^2+(\la_2-\la_1)(\la_4-\la_3)\\
    &\times\sn^2(\sqrt{(\la_4-\la_2)(\la_3-\la_1)}\,\theta,m),
    \end{split}
\end{equation}
\begin{equation}\label{eq17}
\begin{split}
    u(x,t)&=U-{q}/{\rho(x,t)},\\
    q&=-\tfrac18(\la_1-\la_2-\la_3+\la_4)(\la_1-\la_2+\la_3-\la_4)\\
    &\times(\la_1+\la_2-\la_3-\la_4),
    \end{split}
\end{equation}
where
\begin{equation}\label{eq19}
    \theta=x-Ut,\quad U=\tfrac12(\la_1+\la_2+\la_3+\la_4)
\end{equation}
$\nu_1,\nu_2,\nu_3$ are related to the Riemann invariants $\la_i,i=1,2,3,4,$ by the formulae
\begin{equation}\label{eq18}
\begin{split}
    \nu_1&=\tfrac14(\la_1-\la_2-\la_3+\la_4)^2,\\
    \nu_2&=\tfrac14(\la_1-\la_2+\la_3-\la_4)^2,\\
    \nu_3&=\tfrac14(\la_1+\la_2-\la_3-\la_4)^2,
    \end{split}
\end{equation}
and
\begin{equation}\label{eq20}
    m=\frac{\nu_2-\nu_1}{\nu_3-\nu_1}=\frac{(\la_2-\la_1)(\la_4-\la_3)}{(\la_3-\la_1)(\la_4-\la_2)}.
\end{equation}
In the modulated wave the Riemann invariants $\la_i$ become slow functions of $x$ and $t$
and their evolution is governed by the Whitham equations. They were derived for the case of eq.~(\ref{eq1})
with $\Ga=0$ in \cite{kamch2004} and their generalization to the nonzero $\Ga$ is straightforward.
Therefore we will write down here the final result without its derivation.

In our stationary case $\la_i$ do not depend on $t$ and the phase velocity $U$ equals to zero,
\begin{equation}\label{eq21}
    \la_1+\la_2+\la_3+\la_4=0.
\end{equation}
Then the Whitham equations can be written in the form
\begin{equation}\label{eq22}
    \frac{\upd\la_i}{\upd x}=\frac2{L}\cdot\frac{I_1\la_i+I_2}{\prod_{m\neq i}(\la_i-\la_m)},
\end{equation}
where
\begin{equation}\label{eq23}
\begin{split}
    I_1&=\Ga\int_{\nu_1}^{\nu_2}\frac{\nu(\rho_0-\nu)}{\sqrt{\mathcal{R}(\nu)}}\upd\nu,\\
    I_2&=\frac{\Ga u_0\rho_0}2\int_{\nu_1}^{\nu_2}\frac{\rho_0-\nu}{\sqrt{\mathcal{R}(\nu)}}\upd\nu,
    \end{split}
\end{equation}
\begin{equation}\label{eq24}
    \mathcal{R}=(\nu-\nu_1)(\nu-\nu_2)(\nu-\nu_3)
\end{equation}
and $L$ is the wavelength
\begin{equation}\label{eq25}
    L=\frac{2\K(m)}{\sqrt{(\la_3-\la_1)(\la_4-\la_2)}},
\end{equation}
$\K(m)$ being the complete elliptic integral of the first kind.

Due to the special structure of Eqs.~(\ref{eq22}), the symmetric functions of $\la_i$,
\begin{equation}\label{eq26}
\begin{split}
    s_1&=\sum_i\la_i,\quad s_2=\sum_{i\neq j}\la_i\la_j,\quad s_3=\sum_{i\neq j\neq k}\la_i\la_j\la_k,\\
    s_4&=\la_1\la_2\la_3\la_4,
    \end{split}
\end{equation}
obey very simple equations
\begin{equation}\label{eq27}
    \frac{\upd s_1}{\upd x}=0,\quad \frac{\upd s_2}{\upd x}=0,\quad \frac{\upd s_3}{\upd x}=\frac{2I_1}L,\quad
    \frac{\upd s_4}{\upd x}=-\frac{2I_2}L.
\end{equation}
The condensate density $\rho$ oscillates in the interval $\nu_1\leq \rho\leq\nu_2$, i.e. $\nu_1$ and $\nu_2$ are the envelopes
of the oscillations in the dispersive shock wave. Let us study their asymptotic behavior at $x\to-\infty$.

The asymptotic plane wave corresponds to $\nu_2=\nu_1$ (or $\la_2=\la_1$) so that the parameters of the incident condensate
are expressed in terms of the Riemann invariants as
\begin{equation}\label{eq28}
    \rho_0=\tfrac14(\la_4-\la_3)^2
\end{equation}
and
\begin{equation}\label{eq29}
    u_0=\tfrac12(\la_3+\la_4-2\la_1).
\end{equation}
From (\ref{eq21}), (\ref{eq28}) and (\ref{eq29}) we find the asymptotic values of the Riemann invariants
\begin{equation}\label{eq30}
\begin{split}
    \la_1&=\la_2=-\frac{u_0}2,\\
    \la_3&=\frac{u_0}2-\sqrt{\rho_0},\quad \la_4=\frac{u_0}2+\sqrt{\rho_0}.
    \end{split}
\end{equation}
The periodic solution corresponds to the ordering $\la_1\leq\la_2\leq\la_3\leq\la_4$, so that from $\la_3\geq\la_2$
we find that $u_0\geq\sqrt{\rho_0}$, i.e. the incident flow must be supersonic for the generation of such a stationary shock wave.
The wavelength (\ref{eq25}) of small amplitude oscillations ($m=0$) reduces in the limit $x\to-\infty$ to
\begin{equation}\label{eq31}
    L=\frac{\pi}{\sqrt{u_0^2-\rho_0}}.
\end{equation}

The asymptotic values of $\nu_1,\nu_2,\nu_3$ at $x\to-\infty$ are $\nu_1(-\infty)=\nu_2(-\infty)=\rho_0$,
$\nu_3(-\infty)=u_0^2$. We introduce small deviations from these values,
\begin{equation}\label{eq32}
    \nu_1=\rho_0+\delta \nu_1,\quad \nu_2=\rho_0+\delta \nu_2,\quad \nu_3=u_0^2+\delta\nu_3,
\end{equation}
and linearize eqs.~(\ref{eq27}) with respect to $\delta\nu_i$ with account of the identities
\begin{equation}\label{eq33}
\begin{split}
    &\nu_1+\nu_2+\nu_3=-2s_2,\\
    &\nu_1\nu_2+\nu_1\nu_3+\nu_2\nu_3=-s_2^2+4s_4,\quad
    \nu_1\nu_2\nu_3=s_3^2.
    \end{split}
\end{equation}
As a result we arrive at the equations
\begin{equation}\label{eq34}
    \frac{\upd(\delta\nu_1+\delta\nu_2)}{\upd x}=-\frac{\upd\delta\nu_3}{\upd x}=\frac{2\Ga \rho_0u_0}{u_0^2-\rho_0}(\delta\nu_1+\delta\nu_2).
\end{equation}
Hence we get
\begin{equation}\label{eq35}
    \delta\nu_1+\delta\nu_2,\delta\nu_3\propto \exp\left(\frac{2\Ga\rho_0u_0}{u_0^2-\rho_0}x\right).
\end{equation}
These equations suggest that the sum of deviations $\delta\nu_1+\delta\nu_2$ decays at $x\to-\infty$ faster
than the deviations $\delta\nu_1$ and $\delta\nu_2$ separately. If we introduce small deviations of the
Riemann invariants from their asymptotic values (\ref{eq30})
\begin{equation}\label{eq36}
    \begin{split}
    \la_1&=-\frac{u_0}2+\delta\la_1,\quad \la_2=-\frac{u_0}2+\delta\la_2\\
    \la_3&=\frac{u_0}2-\sqrt{\rho_0}+\delta\la_3,\quad \la_4=\frac{u_0}2+\sqrt{\rho_0}+\delta\la_4,
    \end{split}
\end{equation}
then we find that the asymptotic behavior (\ref{eq35}) corresponds to $|\delta\la_3|,|\delta\la_4|\ll|\delta\la_1|,|\delta\la_2|$
so that
\begin{equation}\label{eq37}
\begin{split}
    &\nu_1\cong\tfrac14(2\sqrt{n_0}+\delta\la_1-\delta\la_2)^2,\\
    &\nu_2\cong\tfrac14(2\sqrt{n_0}-\delta\la_1+\delta\la_2)^2,\\
    &\nu_3\cong\tfrac14(2u_0-\delta\la_1-\delta\la_2)^2,
    \end{split}
\end{equation}
and in leading order approximation $\delta\nu_1+\delta\nu_2=0$ which means that (\ref{eq35})
corresponds to higher order corrections.

Thus, in the main approximation we assume that $\nu_3\cong u_0^2$ and from (\ref{eq23}) we obtain the
following equations for $\delta\la_1$ and $\delta\la_2$:
\begin{equation}\label{eq38}
\begin{split}
    \frac{\upd\delta\la_1}{\upd x}&=\frac{\Ga \rho_0u_0}{2(u_0^2-\rho_0)}(\delta\la_1-\delta\la_2),\\
    \frac{\upd\delta\la_2}{\upd x}&=-\frac{\Ga \rho_0u_0}{2(u_0^2-\rho_0)}(\delta\la_1-\delta\la_2).
    \end{split}
\end{equation}
From these equations we get
\begin{equation}\label{eq39}
    \delta\la_{1,2}\cong \pm C\exp\left(\frac{\Ga \rho_0u_0}{u_0^2-\rho_0}x\right)
\end{equation}
($C$ is the integration constant),
and hence the envelopes of oscillations in the dispersive shock decay at leading order approximation as
\begin{equation}\label{eq40}
    \delta\nu_1,\delta\nu_2\propto \exp\left(\frac{\Ga \rho_0u_0}{u_0^2-\rho_0}x\right).
\end{equation}
This is slower than the dependence in (\ref{eq35}) as it should be.

\begin{figure}
\centerline{\includegraphics[scale=0.11]{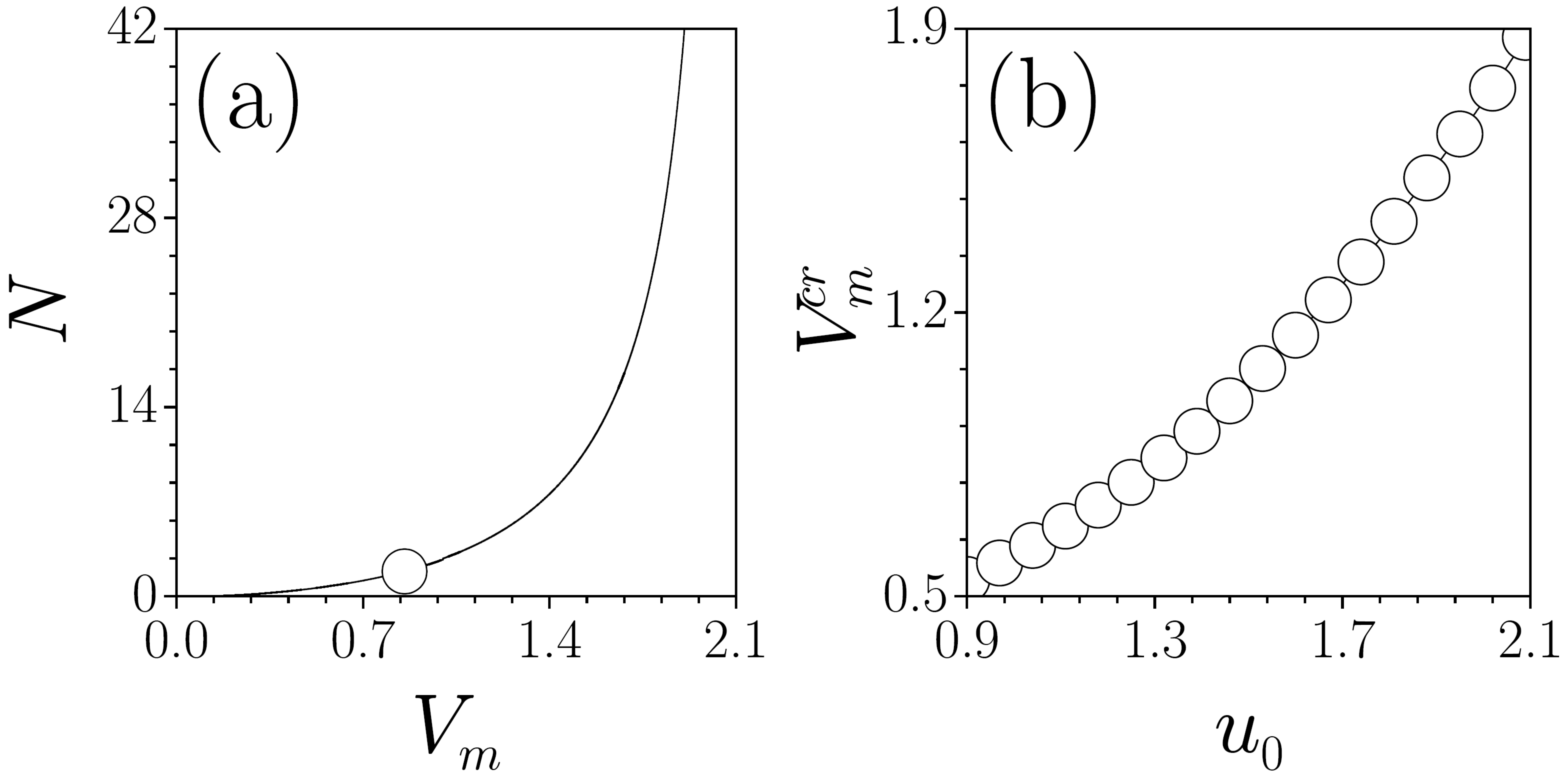}}
\caption{(a) Number of  particles versus potential strength for supersonic  flow with
$u_0=1.4$. The circle corresponds to solution in fig.~\ref{fig.4}.
(b) Critical strength of potential versus $u_0$. In all cases $\ga=\Ga=0.05$.  }
\label{fig.5}
\end{figure}

These analytical predictions were confirmed by the direct solution of eq.~(\ref{eq1}).
As one can see from fig.~\ref{fig.4}, a typical shock wave solution may extend far beyond
the repulsive potential, both in the positive and negative $x$ directions. Note the excellent
agreement at $x\to\pm\infty$ between the numerical density and its analytical fit that uses
the expressions (\ref{eq14}) and (\ref{eq40}). The wavelength of small amplitude oscillations
estimated numerically
for the parameters of the flow shown in fig.~\ref{fig.4} is equal to $L_{num}\cong 3.16$ and
the analytical formula (\ref{eq31}) gives $L_{theor}\cong 3.21$, with an accuracy better
than 2\%.
It should be stressed that the solution in fig.~\ref{fig.4} is fully stationary (that is, it does not change in time),
in contrast to previously reported one-dimensional shock waves in conservative systems. The oscillating
left tail of the dispersive shock wave in the supersonic regime becomes more pronounced with
increase of potential strength $V_m$ and incident velocity $u_0$. Just as for the subsonic flow
the increase of $V_m$ is accompanied by an increase of the amplitude of the shock wave.
However, in the supersonic regime the amplitude of shock wave may be comparable with the amplitude
of the unperturbed plane wave and the density may decrease almost to zero in the downstream region,
especially for strong potentials. The number of particles in the shock wave increases with
$V_m$  [fig.~\ref{fig.5}(a)] and the character of this dependence also points to the existence
of a critical defect strength beyond which shock waves do not exist.

A linear stability analysis performed for stationary solutions shows that wave patterns
are stable for any strength of potential $V_m$
up to the critical one as long as the incident velocity is small enough,
$u_0<0.9c_s$. When $u_0>0.9c_s$, stationary solutions become unstable if the strength
of potential exceeds certain limiting value $V_m^{cr}$ (the instability of
the dispersive shock waves at $V_m>V_m^{cr}$ is accompanied by the development and
emission of small-scale disturbances on the right tail of the shock wave).
Still, even shock waves with very long oscillating tails, like the one shown in fig.\ref{fig.4},
may be stable. The critical value of potential increases monotonically with $u_0$ as
it is shown in fig.~\ref{fig.5}(b).

\section{Conclusion}
It is instructive to compare our results with those obtained in the case of the flow of a conservative
fluid past an obstacle. As was shown in \cite{legk},
in the conservative case one
can distinguish three characteristic ranges of flow velocity---subcritical
($u_0<u_-$), transcritical ($u_-<u_0<u_+$), and supercritical ($u_0>u_+$)
where the critical values of the velocity are located at opposite sides
of the value of the sound velocity ($u_-<c_s<u_+$). If the flow velocity
is subcritical or supercritical, then the disturbance is stationary and has
the dimension of the obstacle's size with definite sign of the difference $\rho-\rho_0$; it has a form
of a dip in subcritical region and of a hump in supercritical region.
In the present case of non-conservative dynamics
the disturbance must have ``dips'' and ``humps'' to be compensated
in the integral (\ref{eq12}). Examples
of such disturbances shown in figs.~\ref{fig.2} and \ref{fig.4}  demonstrate
this property.

In the conservative situation the transcritical regime
corresponds to non-stationary generation of upstream and/or downstream dispersive
shock waves. In the non-conservative situation this regime actually disappears and
instead a supersonic flow generates a {\it stationary dispersive shock wave}
upstream the obstacle. There are no downstream dispersive shock waves; instead they are
replaced by smooth profiles decaying asymptotically to the stationary plane
wave.

The above properties of the wave patterns generated by the flow of polariton
condensate past an obstacle are derived for
the model (\ref{eq1}).
We believe, however, that these properties remain qualitatively
the same for other models having stationary plane wave states formed due to balance
of pumping and dissipation effects. In particular, this theory can find applications
to a so-called ``superfluid motion of light'' \cite{pr-1993,bcz-2001,lm-2010}.

\acknowledgments
We are grateful to A.~Amo, N.~Berloff, J.~Bloch,
A.~Bramati, I.~Carusotto, C.~Ciuti, E.~Giacobino, Yu.G.~Gladush, N.~Pavloff, D.~Sanvitto
for discussions of superfluidity in the cavity polariton physics.
We thank RFBR for partial support.

\end{document}